\documentclass[referee]{aa}

\usepackage{graphicx}
\usepackage{txfonts}
\usepackage{natbib}
\bibpunct{(}{)}{;}{a}{}{,}

\newcommand\about  {\hbox{$\sim$}}
\renewcommand\deg  {\hbox{$^\circ$}}
\newcommand{\E}[1]{\hbox{$10^{#1}$}}
\newcommand\x      {\hbox{$\times$}}
\newcommand\Lo     {\hbox{$L_{\odot}$}}
\newcommand\Mo     {\hbox{$M_{\odot}$}}

\newcommand\kms   {\hbox{km\,s$^{-1}$}}

\begin{document}

\title{H$_2$D$^+$ line emission in Proto-Planetary Disks}
\author{Andr\'es Asensio Ramos\inst{1}
  \and Cecilia Ceccarelli\inst{2}
  \and Moshe Elitzur\inst{3}}

\offprints{Cecilia.Ceccarelli@obs.ujf-grenoble.fr}

\institute{
 Instituto de Astrof\'{\i}sica de Canarias, 38205, La Laguna, Tenerife, Spain
 \and
 Laboratoire d'Astrophysique de l'Observatoire de Grenoble,
BP 53, 38041 Grenoble, Cedex 9, France
 \and
 Physics \& Astronomy Department, University of Kentucky, Lexington, KY
 40506-0055, USA
\\
\email{aasensio@iac.es, Cecilia.Ceccarelli@obs.ujf-grenoble.fr,
moshe@pa.uky.edu}
}
\date{Received ; Accepted}
\titlerunning{H$_2$D$^+$ line profiles}
\authorrunning{Asensio Ramos et al.}

\abstract
{Previous studies have indicated that the 372.4 GHz ground transition of
ortho-H$_2$D$^+$ might be a powerful probe of Proto-Planetary Disks. The line
could be especially suited for study of the disk mid-plane, where the bulk of
the mass resides and where planet formation takes place.
}
{Provide detailed theoretical predictions for the line intensity, profile and
maps expected for representative disk models.
}
{We determine the physical and chemical structure of the disks from the model
developed by Ceccarelli \& Dominik (2005). The line emission is computed with
the new radiative transfer method developed recently by Elitzur \& Asensio
Ramos (2006).
}
{We present intensity maps convolved with the expected ALMA resolution, which
delineate the origin of the H$_2$D$^+$ 372.4 GHz line. In the disk inner
regions, the line probes the conditions in the mid-plane out to radial
distances of a few tens of AU, where Solar-like planetary systems might form.
In the disk outermost regions, the line originates from slightly above the
mid-plane. When the disk is spatially resolved, the variation of line profile
across the image provides important information about the velocity field.
Spectral profiles of the entire disk flux show a double peak shape at most
inclination angles.
}
{Our study confirms that the 372.4 GHz H$_2$D$^+$ line provides powerful
diagnostics of the mid-plane of Proto-Planetary Disks. Current submillimeter
telescopes are capable of observing this line, though with some difficulties.
The future ALMA interferometer will have the sensitivity to observe and even
spatially resolve the H$_2$D$^+$ line emission.
}


\maketitle


\section{Introduction}\label{sec:introduction}

Proto-Planetary Disks are believed to mark the transition between the
protostellar phase and planet formation. Similar to ``standard'' molecular
clouds, young Proto-Planetary Disks are gas rich, with gas to dust mass ratio
of \about 100, and their dust grains have average sizes of \about 0.1 $\mu$m.
As the disk evolves with time, the dust grains coagulate into progressively
larger bodies --- the seeds for planet formation. The gas is dispersed via
viscous accretion and/or evaporation \citep[e.g.][]{2004ApJ...611..360A}.
Following the details of this evolution is of paramount importance for
understanding the process of planet formation. Especially relevant in this
context is the disk mid-plane because it contains the bulk of the mass and
because its magneto-rotational instabilities are believed to be a major source
of the disk viscous accretion \citep{1996ApJ...462..725G}; thus determining the
degree of ionization is also of great importance \citep[see
also][]{2004A&A...417...93S}.

Observations of the 372 GHz $1_{1,0}-1_{1,1}$ ground-state transition of
ortho-H$_2$D$^+$ provide the best, and possibly only, current means for
studying these aspects of {\it disks around solar type protostars}. The reason
is that in the disk mid-plane the temperature is so low ($\leq 20$ K) and the
density so high ($\geq \mathrm{1\times10^{6}}$ cm$^{-3}$) that virtually all
heavy-elements bearing molecules freeze-out onto the grain mantles,
disappearing from the gas phase. Only molecules containing H and D atoms stay
gaseous and are capable of probing the gas in the disk mid-plane. Because the
ground transitions of H$_2$ and HD, the most abundant H and D bearing
molecules, have relatively high energies ($\geq 100$ K), they do not get
excited in that cold environment. In principle, these transitions could be seen
in absorption against the dust continuum, but such observations are extremely
difficult because, if for no other reason, the relevant wavelengths are blocked
by the Earth atmosphere (28 and 112 $\mu$m for H$_2$ and HD, respectively)
\footnote{Previous
  detections of the H$_2$ S(0) and S(1) transitions towards
  Proto-Planetary Disks by \cite{2001Natur.409...60T} have not been
  confirmed \citep{2005ApJ...620..347S,2006ApJ...651.1177P}.}.  As a
result, the most promising probes of the disk mid-plane are
transitions of the next most abundant molecules containing H and D
atoms: H$_3^+$, formed by the interaction of cosmic rays with H$_2$
and H, and its deuterated forms H$_2$D$^+$, HD$_2^+$ and
D$_3^+$. While the symmetric molecules H$_3^+$ and D$_3^+$ do not
possess dipole rotational transitions, both H$_2$D$^+$ and HD$_2^+$
have ground transitions in the sub-millimeter region that can, in
principle, be observed with ground based telescopes. Moreover, the
disappearance of heavy-element bearing molecules from the gas phase
enhances the H$_2$D$^+$/H$_3^+$ and HD$_2^+$/H$_3^+$ ratios to values
that can exceed unity \citep{2003ApJ...591L..41R} --- H$_2$D$^+$ and
HD$_2^+$ become the most abundant positive charge carriers under these
circumstance, and thus are not only the best probes of the disk
mid-plane but also of the ionization degree.  Detailed models of the
chemistry of Proto-Planetary Disks confirm and quantitatively support
these general arguments
\citep{2005A&A...440..583C,2007astro.ph..1484W}.

Both H$_2$D$^+$ and HD$_2^+$ come in ortho and para forms, depending
on the spin alignment of the H and D pairs. Of the five possible
ground state transitions of the four species, only three have been
observed thus far: the ortho-H$_2$D$^+$ $1_{1,0}-1_{1,1}$ at 372.4 GHz
\citep{1999ApJ...521L..67S} and the $2_{1,2}-1_{1,1}$ at 2363 GHz
(Cernicharo et al. 2007), and the para-HD$_2^+$ $1_{1,0}-1_{0,1}$ at
691.7 GHz \citep{2004ApJ...606L.127V}.  The two other transitions lie
in a frequency domain inaccessible from the ground (at 1370.1 and
1476.6 GHz) and have not been detected thus far; this should change
with the anticipated launch of Herschel, followed by SOFIA.
Presently, the para-HD$_2^+$ line has been detected in only one
Pre-Stellar Core (Vastel et al. 2004). In contrast, the
ortho-H$_2$D$^+$ line at 372 GHz has been observed in several objects,
mostly Pre-Stellar Cores, where the conditions are similar to those in
the disk mid-plane
\citep{2003A&A...403L..37C,2006A&A...454L..59H,2006A&A...454L..55H},
and in one Proto-Planetary Disk \citep{2004ApJ...607L..51C}. In
Pre-Stellar Cores, the line is so bright that both the line profile
\citep{2005A&A...439..195V} and the emission spatial extent
\citep{2006ApJ...645.1198V} have been studied.  Unfortunately, the
compact dimensions of Proto-Planetary Disks preclude such studies
there because of the limited sensitivity and spatial resolution of
current instruments. However, the future Atacama Large Millimeter
Array (ALMA) will have the capacity to not only easily detect the
H$_2$D$^+$ line in disks but also to carry out mapping and line
profile measurements, similar to what has already been done toward
Pre-Stellar Cores. These studies will provide us with maps of the
ionization structure in the disk mid-plane, and the dynamics;
possibly, the presence of forming planets might be detected in
perturbations of Keplerian motions.

In anticipation of this progress in observational capabilities, here we report
a study of the ortho-H$_2$D$^+$ $1_{1,0}-1_{1,1}$ line emission in
Proto-Planetary Disks. We present theoretical predictions for the line profiles
and the line emission maps in a variety of Proto-Planetary Disk models. The
calculations build on the study of the chemistry in the outer mid-plane of
Proto-Planetary Disks by \cite{2005A&A...440..583C}, and employ the new
radiative transfer formalism recently developed by \cite{2006MNRAS.365..779E}
to compute the H$_2$D$^+$ line emission. Details of the modeling procedure,
including non LTE radiative transfer calculations and chemical/physical models,
are described in \S\ref{sec:model}. The results of our computations and
implications for observability of the line and its diagnostic capability are
described in \S\ref{sec:results}. Our conclusions are contained in
\S\ref{sec:concl} .

\section{The model}\label{sec:model}

\subsection{The disk model}\label{sec:diskmod}

Following the strategy described in \cite{2005A&A...440..583C}, we use
the grid of models computed in that study. We consider a
Proto-Planetary Disk in Keplerian rotation around a protostar with
mass 0.5 \Mo, luminosity 0.5 \Lo\ and T$_{\rm eff}$ = 3630 K. Starting
at an inner radius of 45 AU, the disk surface density follows a power
law $\Sigma \propto r^{-1}$, namely, the mass per unit radius is
constant across the disk. The gas is mixed with the dust, and we fix
the mass of the disk {\it dust} content at $2\times10^{-4}$
M$_\odot$. Model number 1 corresponds to a set of ``standard''
parameters: disk outer radius of 400 AU; dust-to-gas mass ratio of
1:100, i.e., the disk mass is 0.02 M$_\odot$; average grain radius of
0.1 $\mu$m; and cosmic ray ionization rate $\zeta_{\rm cr} =
3\times10^{-17}$ s$^{-1}$\footnote{We assume a constant cosmic rays
  ionization rate across the disk, as they can penetrate up to a
  column density of $\sim 100$ gr cm$^{-2}$, never exceeded in the
  disk regions relevant to the present study..}. We consider four
additional models, varying these parameters one at a time in each
case. In model 2 the grain radius is increased to 1 $\mu$m,
corresponding to a case in which a substantial fraction of the dust
has coagulated so that the small grains are removed. Model 3 checks
the effect of increasing the cosmic ray ionization rate by factor ten,
as might be the case of either a larger $\zeta$ or when emulating
strong X-ray irradiation of the disk. In model 4 the dust-to-gas ratio
is increased by factor ten, so that the disk mass is only 2\x\E{-3}
M$_\odot$. This corresponds to a phase in which a fair amount of gas
has been dispersed. Finally, model 5 corresponds to a larger disk
radius of 800 AU, similar to the well studied disk of DM Tau, where
H$_2$D$^+$ has been detected \citep{2004ApJ...607L..51C}. Table 1
summarizes the parameters of all models.

\begin{table}[ht]
  \centering
  \begin{tabular}{|c|cccc|}
    \hline
    \hline
    Model & dust:gas   & $\zeta_{\rm cr}$ & $a_{\rm grain}$ & disk radius\\
          & mass ratio & [$3\times10^{-17}$ s$^{-1}$] & [$\mu$m] & [AU]\\
          \hline
    1     & 0.01   & 1 & 0.1 & 400\\
    2     & 0.01   & 1 & 1.0 & 400\\
    3     & 0.01   & 10& 0.1 & 400\\
    4     & 0.1    & 1 & 0.1 & 400\\
    5     & 0.01   & 1 & 0.1 & 800\\
    \hline
    \hline
  \end{tabular}
\caption{The five disk models considered in this study; $\zeta_{\rm cr}$ is the
cosmic rays ionization rate, $a_{\rm grain}$ is the average grain radius.
}
  \label{tab:models}
\end{table}

\begin{figure*}[ht]
  \centering
  \includegraphics[angle=0,width=1.\columnwidth]{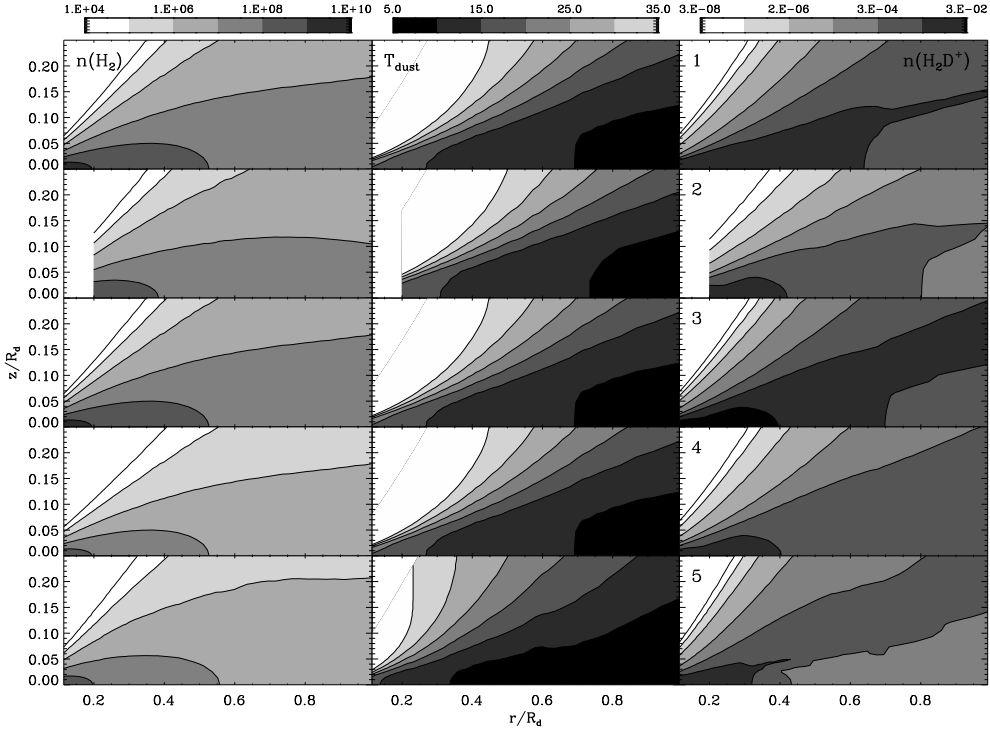}
\caption{Physical and chemical structure of the models listed in Table 1; each
row corresponds to the model number listed in its right panel. Each panel
presents a contour map whose axes denote radial and vertical distances
normalized to the disk radius (note that $R_{\rm d}$ is doubled in Model 5).
{\em Left column}: H$_2$ density. {\em Center column}: Dust temperature. {\em
Right column}: H$_2$D$^+$ density. Contour units are cm$^{-3}$ for the H$_2$
and H$_2$D$^+$ densities, and K for the dust temperature. The contour values
are denoted in the bar on top of each column. Note that Model 2 is truncated at
$r/R_{\rm d}$ = 0.2, where the CO depletion is larger than 3; the Ceccarelli \&
Dominik (2005) model is no longer adequate in this regime.
}
 \label{fig:models}
\end{figure*}

Using the approach described in \cite{2001ApJ...560..957D} and
\cite{2004A&A...417..159D} for a passively irradiated hydrostatic flaring
structure, the disk physical structure is derived self-consistently. {\bf The
gas temperature is assumed to be the same as the dust. This is a justified
assumption for the regions relevant to the present study, where the density is
larger than about $10^{5}$ cm$^{-3}$ and the gas and dust are thermally coupled
because of the gas-dust collisions (e.g. Goldsmith 2001). We also assume a
constant dust-to-gas ratio throughout the disk, assuming that no sedimentation
of dust occurred yet.} Fig. \ref{fig:models} shows the radial and vertical
profiles of the gas density and the dust temperature for the 5 models
considered here. The chemical composition, in particular the H$_2$D$^+$ radial
and vertical abundance profiles, is computed by solving for chemical
equilibrium, as described in \cite{2005A&A...440..583C}, and is shown in Fig.
\ref{fig:models} for the 5 models. The H$_2$D$^+$ abundance depends primarily
on: i) the cosmic rays ionization rate, which governs the overall ionization
structure; ii) the dust-to-gas ratio, which determines the disk mass; and iii)
the grain average radius $a_{\rm grain}$. The grain size affects the overall
ionization structure since recombination with negatively charged grains can
become the major destruction channel for H$_3^+$, H$_2$D$^+$ etc. In addition,
$a_{\rm grain}$ affects also the deuteration ratio (i.e. H$_2$D$^+$/H$_3^+$)
since it controls the surface area on which heavy-elements bearing molecules
freeze-out. The models considered here explore the influence of each relevant
parameter on the H$_2$D$^+$ $1_{1,0}-1_{0,1}$ line. Based on the work by
\cite{2005A&A...440..583C}, this ensemble of models should provide a good
representation of the range of physical conditions relevant to H$_2$D$^+$ line
emission in Proto-Planetary Disks.

The H$_2$D$^+$ density, which is controlled primarily by the depletion of
heavy-elements bearing molecules due to freeze-out, is expected to peak at
``intermediate'' depletion --- at low depletion H$_3^+$ becomes more abundant
than H$_2$D$^+$, and at very large depletion D$_3^+$ becomes the most abundant
positive charge carrier, as it is the end of the chain transferring D atoms
from HD. Indeed, the ``standard'' Model 1, displayed in the top row of figure
\ref{fig:models}, shows that the H$_2$D$^+$ density is largest in the disk
mid-plane, where the depletion is ``intermediate'', as expected. The
``intermediate depletion'' region shrinks when the grain size increases (Model
2, second row in the figure) and expands when the cosmic ionization rate
increases  (Model 3, third row); furthermore, the peak H$_2$D$^+$ density is
higher in the latter case. Increasing the dust-to-gas ratio implies a lower gas
density (as the disk dust mass is fixed), and a comparatively reduced region of
H$_2$D$^+$ at a given density (Model 4). Finally, keeping the disk mass fixed
while increasing its radius (Model 5) reduces the overall density and decreases
the portion of the disk where H$_2$D$^+$ is the most abundant; most of the
outer disk mid-plane is now so cold that the major charge carrier there is
D$_3^+$.

\subsection{The radiative transfer calculation}\label{sec:radtrans}

At every point in the disk, the populations of the two levels of the
$1_{1,0}-1_{1,1}$ transition are determined from the balance between
excitations and de-excitations due to collisional and radiative interactions.
Given the population distribution, the intensity emerging along any line of
sight can be obtained from ray tracing. However, the radiative interactions
involve the local radiation field, which includes the contribution of line
emission from the entire disk (the ``diffuse radiation'') and thus cannot be
determined before the level populations are known everywhere. Determining the
level populations requires coupling between the radiative transfer and
statistical rate equations. In the recently developed Coupled Escape
Probability (CEP) method, the effects of the diffuse radiation are incorporated
into the level population equations through coefficients that rigorously
account for the radiative coupling between different regions
\citep{2006MNRAS.365..779E}. The exact solution of the problem is obtained from
a set of non-linear algebraic equations for the level populations in each zone.
In a comparative study of a number of standard problems, the CEP method
outperformed the leading accelerated $\Lambda$-iteration by substantial
margins.

With a grid of 80 radial points and 100 vertical points we divide the disk into
8000 rings. Each ring has a rectangular cross-section and uniform physical
conditions. Thus far the CEP method has been formulated only for the slab
geometry, therefore its application here involves an approximation. Consider
the vertical profiles of density, etc., in the disk at a given radius, and a
stratified slab with the same structure perpendicular to its surface.
Neglecting radiative energy flow in the radial direction through the disk, the
vertical profile of the level population distribution at that radius can be
obtained from the solution of the stratified slab. This is the approach we take
here. The approximation is justified because radiative coupling is suppressed
in the radial direction not only by the large optical depths but also by the
Doppler shifts due to differential Keplerian rotation.

Following the suggestion by Stark et al.\ (1999) that the collisional
de-excitation rate coefficient is $\sim 10^{-10}$ cm$^3$s$^{-1}$, the
transition critical density used here is $1\times10^{6}$ cm$^{-3}$. New
unpublished computations (E. Hugo \& S. Schlemmer, private communication)
suggest a rate coefficient of $\sim10^{-9}$ cm$^3$s$^{-1}$ at 10 K; changing
the temperature to 20 K has only a 5\% effect on the result in these
computations. The implied critical density is a factor 10 lower than assumed
here. Since the densities of the emitting gas are larger than either value, the
transition level populations are practically always in thermal equilibrium so
that the exact value of the collision rate is not critical. {\bf Finally, we
assume a constant ortho-to-para H$_2$D$^+$ ratio across the disk equal to 0.3,
as in Ceccarelli \& Dominik (2005).}

\section{Results}\label{sec:results}

The aim of this study is to provide predictions for comparison with actual
observations of the ortho-H$_2$D$^+$ ground line.  Two categories of
measurements exist --- single-dish and interferometric observations. We present
the results of our calculations for these two cases separately in each of the
following subsections. In computing the line emission, we assume the H$_2$D$^+$
ortho-to-para ratio is 0.3 \citep{2005A&A...440..583C} and consider a source at
a distance of 140 pc.

\subsection{Spectral line profiles}\label{sec:standard}

\begin{figure*}[tb]
  \centering
 \includegraphics[angle=0,width=1.\columnwidth]{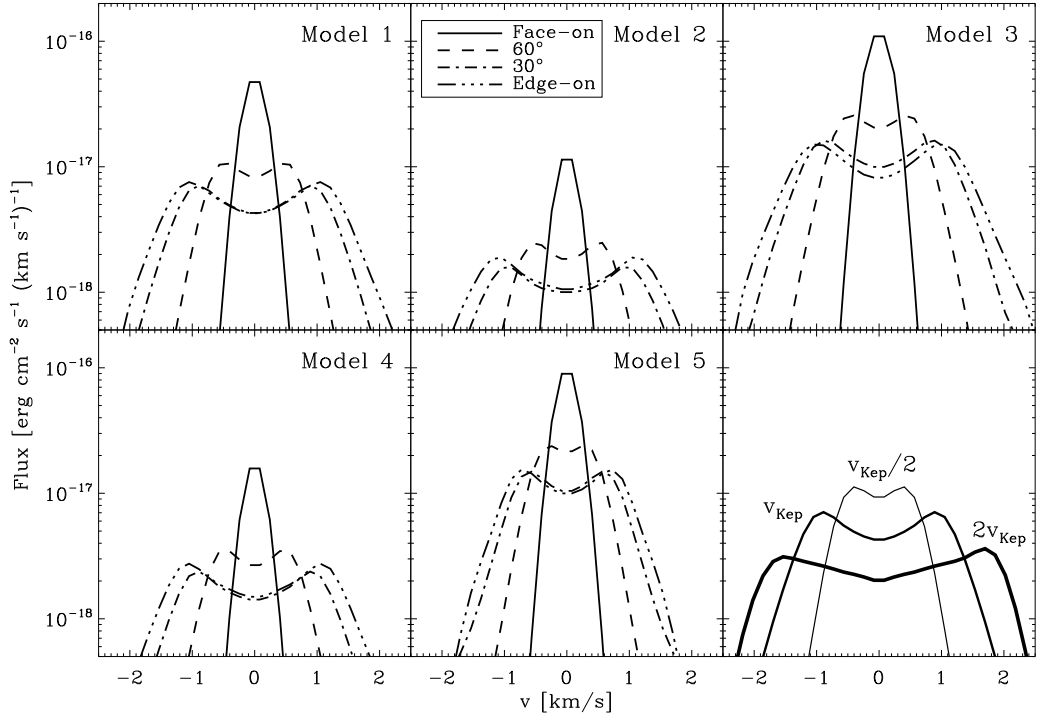}
\caption{Spectral line profiles of the disk flux at a distance of 140 pc for
each model at various viewing angles, as marked (face-on is 90\deg). For
demonstrating the effects of rotation, the bottom-right panel reproduces the
profile of Model 1 at 30\deg\ inclination together with the same calculation in
which the Keplerian rotation is speeded up and down by factor 2, as marked.
}
  \label{fig:profiles}
\end{figure*}

At present, few single-dish telescopes are capable of observing the H$_2$D$^+$
372 GHz line: the 15m JCMT, 12m APEX, 10m CSO and 3m KOSMA. At a distance of
140 pc, a disk radius of 400 AU corresponds to an angular diameter of \about
6\arcsec, less than the smallest telescope beam currently available (JCMT with
15\arcsec\ at 372 GHz). The only measurable quantity is the overall flux
emerging from the disk. Figure \ref{fig:profiles} shows the flux spectral shape
for the five models listed in Table 1 for different viewing angles, from
face-on to edge-on. The flux level is highest in Model 3, which has the highest
H$_2$D$^+$ density (see figure \ref{fig:models}) owing to its increased cosmic
ray ionization rate, and in model 5, which has the largest surface area. It is
lowest in Models 2 and 4, where the H$_2$D$^+$ density is reduced.

All models show a similar behavior for the line profiles. Face-on viewing
produces single-peak profiles with the thermal line width. As the disk
orientation changes toward edge-on, the profiles broaden and switch to a
double-peak shape reflecting the disk rotation. The peak separation is
determined by the combined effects of column density and velocity projection
along the line of sight. For the 400 AU disks, the Keplerian velocity varies
from 3.2 \kms\ at 45 AU to 1.1 \kms\ at 400 AU, with the largest line optical
depth in the vertical direction (with a value of \about 0.5 for Model 1)
occurring at a radius of \about\ 220 AU, where the velocity is 1.4 \kms.  To
demonstrate the effect of rotation speed on the profile shape, figure
\ref{fig:profiles} shows also profiles in which the disk rotation is
arbitrarily speeded up or slowed down by factor 2 while all other model
parameters are held fixed. Although the profiles change their shape with
inclination, the area under each curve remains almost the same; that is, the
velocity-integrated flux is roughly independent of viewing angle. The computed
values of the velocity-integrated flux agree rather well with the estimates
obtained by \cite{2005A&A...440..583C} with the standard escape probability
approximation.

 \subsection{Line optical depth}
\begin{figure*}[tb]
  \centering
  \includegraphics[angle=0,width=1.\columnwidth]{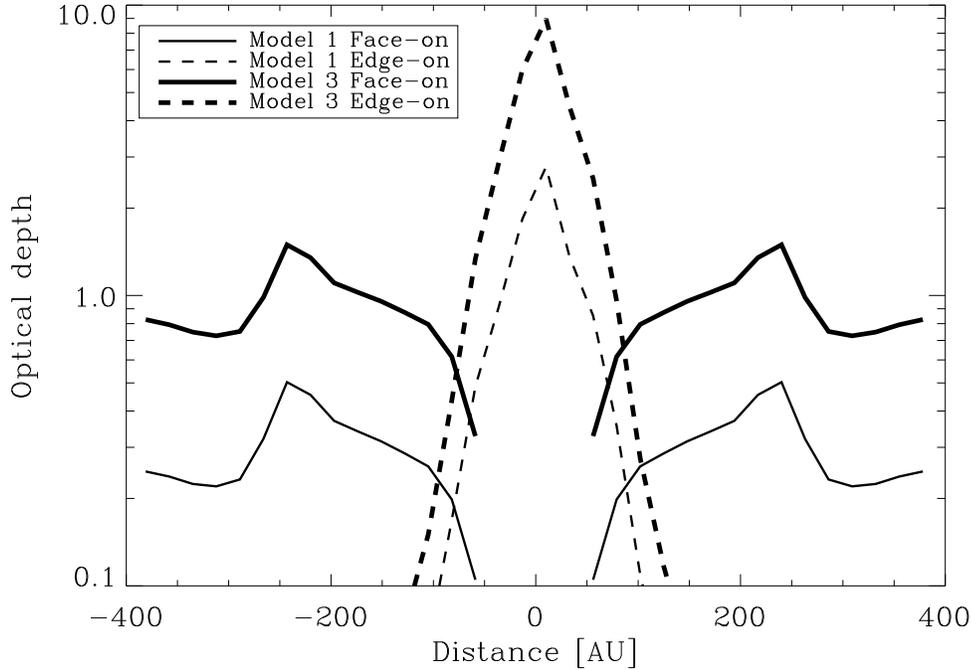}
  \caption{Optical depths at the core of the line for face-on and
    edge-on orientations. Shown are cuts along axes passing through
    the center of the disk, so that the optical depth is maximized.
    The results for model 3 represent an upper limit for the optical
    depth among all models. Model 1 presents optical depths that are
    intermediate between model 3 and all other models. {\bf Note that
      the zero optical depth in the central region of the
      face-on disk is a numerical artifact, as our model does not
      consider that region.} }
  \label{fig:optical-depth}
\end{figure*}

It is of interest to investigate the line optical depth for different models
and different positions. Figure \ref{fig:optical-depth} shows the optical depth
at the core of the line for models 1 and 3 for cuts passing through the disk
center in the two limiting orientations of face-on and edge-on. Thanks to an
increased cosmic ray ionization rate that leads to a higher H$_2$D$^+$ density,
the results of model 3 represent an upper limit on the optical depths.  In
edge-on orientation the inner 100 AU region is optically thick while the
face-on case is neither optically thick nor thin, except for a ring at
$\sim$250 AU. This ring can be recognized in the images shown in Fig.
\ref{fig:integrated-maps}, even though the inclination angle is different.
Model 1 produces very similar results but with reduced optical depths. In this
case, only the edge-on orientation yields optical depths above unity.

\subsection{Brightness maps}\label{sec:others}

\begin{figure*}[tb]
  \centering
  \includegraphics[angle=0,width=1.\columnwidth]{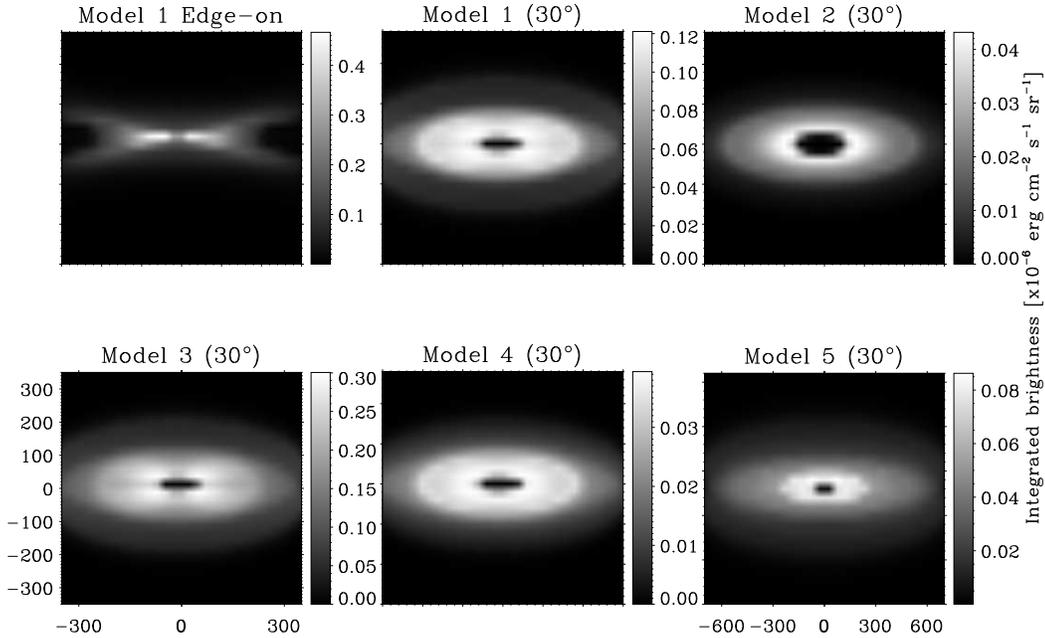}
\caption{Maps averaged over a representative ALMA beam (Gaussian with FWHM =
0.3\arcsec) of the velocity-integrated intensity. The bar to the right of each
panel shows the contour scales. The x and y axes are displacements, in AU, from
the center of the disk. Top left panel is for an edge-on (90\deg) view of Model
1. The other panels show all five disk models (see Table 1) at a viewing angle
of 30\deg. Note the larger spatial extent of Model 5.
}
  \label{fig:integrated-maps}
\end{figure*}

Interferometic observations with SMA (and eSMA), IRAM Plateau de Bure
and ALMA will be able to spatially resolve the H$_2$D$^+$ 372 GHz line
emission from disks like the ones considered here. We produce model
maps by convolving the intensity with a Gaussian beam with FWHM of
0.3\arcsec, corresponding to a linear size of 40 AU at a distance of
140 pc; such resolution can be reasonably expected at ALMA for the
predicted intensities. Fig.\ \ref{fig:integrated-maps} shows some
results. The edge-on map of Model 1 traces closely the horizontal and
vertical distributions of the H$_2$D$^+$ density. The most intense
emission is produced in the inner regions close to the disk mid-plane
within a height of \about\ 1/10 of the radius, confirming the
suggestion that the ortho-H$_2$D$^+$ 372 GHz line probes the disk
mid-plane \citep{2004ApJ...607L..51C}. The high brightness is
maintained out to radii approaching 100 AU, covering the region where
a solar-like planetary system might be forming. Therefore this line
can serve as a probe of the ionization degree and kinematics in a
region critical for planet formation. Further out, because of the
extreme molecular depletion in the equatorial plane, which shifts all
deuterium into D$_3^+$ (see Fig.  \ref{fig:models} and the discussion
in \S \ref{sec:diskmod}), the H$_2$D$^+$ line probes regions slightly
above it.

\begin{figure*}[tb]
  \centering
  \includegraphics[angle=0,width=1.\columnwidth]{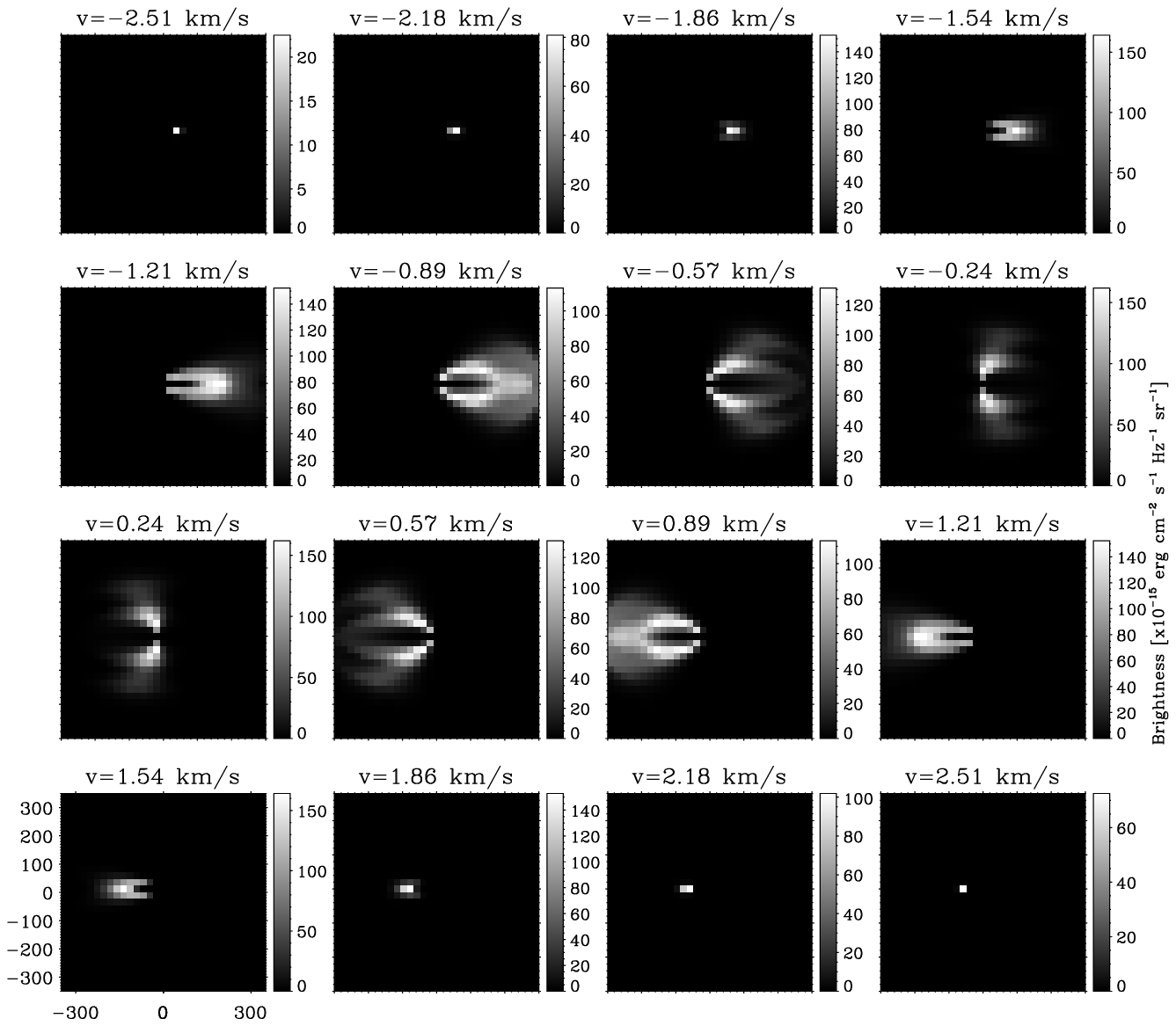}
\caption{Channel maps of the H$_2$D$^+$ line intensity (convolved with ALMA
beam; see Fig. \ref{fig:integrated-maps}) for the standard disk (Model 1),
viewed at an inclination of 30\deg. The x and y axes are displacements, in AU,
from the disk center. The central velocity of each map is marked on top. The
bar to the right of each panel shows the contour brightness scale. Note the
scale changes among the panels. }
  \label{fig:channels}
\end{figure*}

\begin{figure*}[tb]
  \centering
  \includegraphics[angle=0,width=1.\columnwidth]{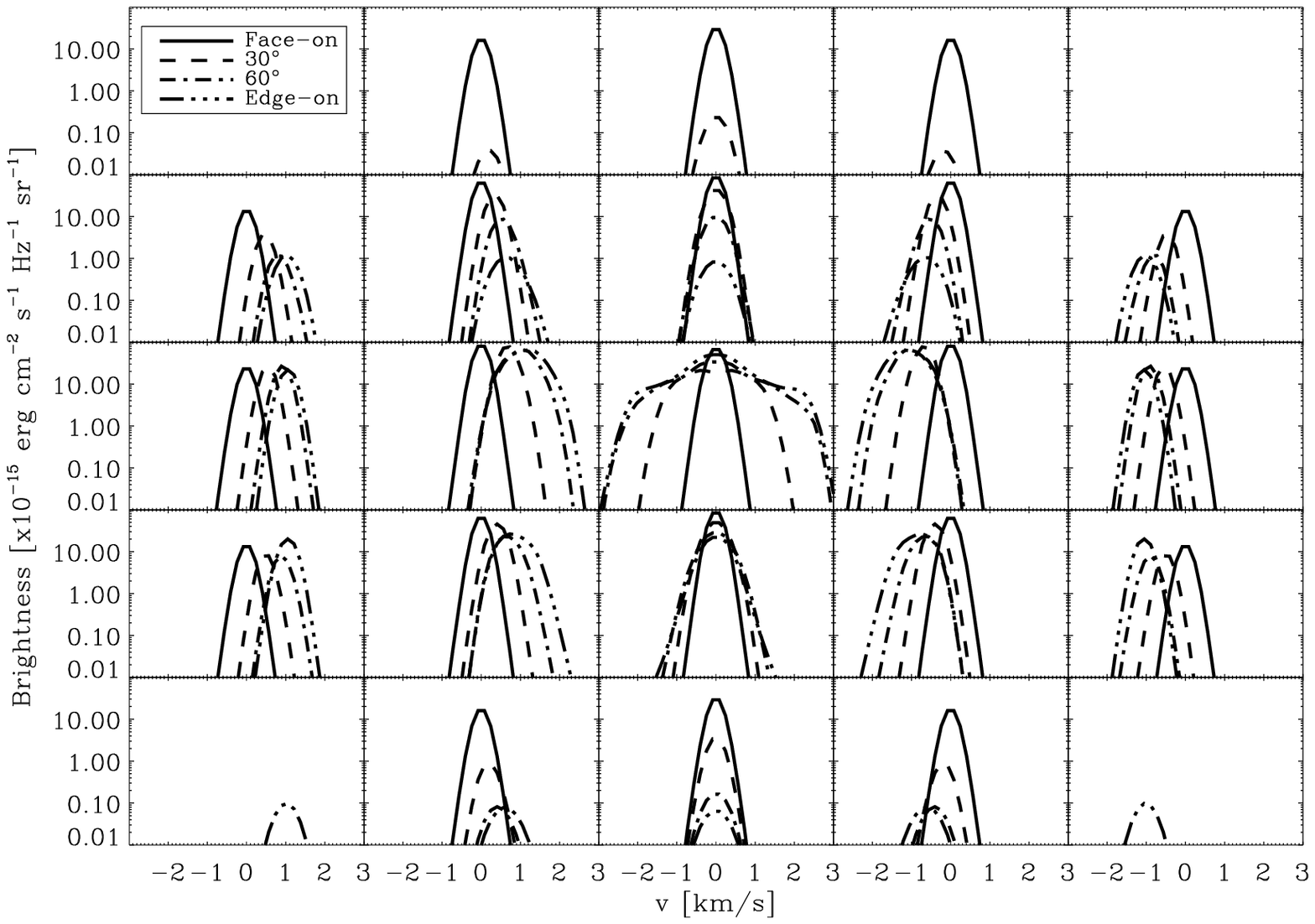}
\caption{Variation of line profile across the disk image for Model 1 at various
inclinations, as marked. The displayed mosaic is centered on the disk image and
was produced as follows: The maps shown in figure \ref{fig:integrated-maps} are
divided into grids of 5\x5 squares, each with 1.2\arcsec\ on the side. At a
given velocity, the brightness is averaged over each square with a Gaussian
beam of 0.3\arcsec, the ALMA beam size.
}
  \label{fig:mosaic-profiles}
\end{figure*}

The five model maps at 30\deg\ inclination show additional structure,
reflecting variation in column density and path length due to the geometry and
chemistry of the flaring disk. These variations control also the channel maps,
shown in Fig. \ref{fig:channels} for the standard disk (Model 1) at 30\deg\
inclination. Brighter and fainter rings are evident at different positions for
different velocities, depending on the intercepted portion of disk. Overall,
the maps show the standard Keplerian pattern with blue-shifted and red-shifted
lobes spatially separated and symmetric with respect to the rotation axis. Fig.
\ref{fig:mosaic-profiles} shows a mosaic of the line profiles for this case.

\textbf{All maps employ intensity units. Those can be converted to ``equivalent
Rayleigh-Jeans temperature'' through $T_{\rm RJ} = 2.35\,I_{-13}$ K, where
$I_{-13}$ is the intensity in units of
\E{-13}~erg~cm$^{-2}$~s$^{-1}$~Hz$^{-1}$~sr$^{-1}$. It should be noted that,
since the temperature equivalent of the 372 GHz transition frequency (806
$\mu$m wavelength) is 18 K, the actual brightness temperature is usually quite
different from $T_{\rm RJ}$.
}

\section{Conclusions}
\label{sec:concl}

Our results confirm the utility of the 372 GHz H$_2$D$^+$ line in probing the
inner regions of Proto-Planetary Disks. While the predicted line fluxes are at
the limit of detectable power of current sub-millimeter telescopes, the future
ALMA interferometer will be able to not only detect the line but also study the
distribution and extent of the H$_2$D$^+$ emission. The spatially and velocity
integrated intensities computed in this study (\S\ref{sec:results}) compare
rather well with the previous estimates by \cite{2005A&A...440..583C}, which
were derived in the escape probability approximation. Indeed,
\cite{2006MNRAS.365..779E} note that the escape probability approximation is
adequate for calculating the line integrated emission of a two-level system
from a homogeneous slab. Our study suggests that this approximation can be used
to provide rough estimates of the integrated line intensity also in the case of
disks with variable physical properties. However, line profile calculations
require an exact formalism, like the CEP method employed here.

The major uncertainty in our modeling is the disk physical structure and
H$_2$D$^+$ abundance. As pointed out by \cite{2005A&A...440..583C}, one major
uncertainty is linked to the role of N$_2$ and, specifically, its freezing-out
onto the grain mantles. The models studied here adopt a sticking coefficient of
1 and a binding energy of 575 K, as suggested by early models and observations
\citep[e.g.][]{2002ApJ...570L.101B}. However, laboratory studies show that the
binding energy of N$_2$ is similar to that of CO, namely 885 K. If this is the
case, N$_2$ will freeze-out more easily and a larger region of the disk than
computed here would be dominated by D$_3^+$, at the expense of H$_2$D$^+$. The
resulting H$_2$D$^+$ column density would then be a factor 30 lower for the
standard case (Model 1), making the line much more difficult to detect
\citep[see also ][]{2007astro.ph..1484W}. Complementary observations of
N-bearing molecules, like N$_2$H$^+$, will be necessary for drawing a coherent
picture. A second uncertainty involves the H$_2$D$^+$ ortho-to-para ratio,
which is observationally entirely unknown. Theoretical estimates vary depending
on several parameters, most of them unknown too, like the ortho-to-para ratio
of H$_2$ and the grain size
\citep{2004A&A...427..887F,2006ApJ...645.1198V,2006A&A...449..621F}. Clarifying
this issue requires observations of the ground transition of the
para-H$_2$D$^+$ at 1370.1 GHz. Unfortunately, no current facility is capable of
such observations, and neither will Herschel.  A third uncertainty arises from
the collision rate coefficient of the $1_{1,1,}-1_{1,0}$ transition, but this
should have only a minor effect given the relatively high densities in the disk
mid-plane.

With all of these caveats and in spite of the uncertainties, observations of
H$_2$D$^+$ at 372 GHz remain a powerful diagnostic tool for probing the disk
mid-plane. Imaging of the 372 GHz line emission will bring information,
otherwise unavailable, on the disk kinematics and ionization structure, serving
as a unique tool in the study of planet formation. For example, ALMA will be
able to reach a rms of $\sim$2 mJy on a 0.3 km/s velocity bin and 0.3$"$ beam
in 10 hr of integration time\footnote{based on the ALMA time estimator;
http://www.eso.org/projects/alma/science/bin/sensitivity.html}, corresponding
to a line brightness of $1\times10^{-14}$
erg~cm$^{-2}$\,s$^{-1}\,$Hz$^{-1}$\,sr$^{-1}$. Comparison with the predictions
reported in Fig. \ref{fig:channels} shows that this will allow imaging of the
H$_2$D$^+$ line emission at the scale expected for planet formation, providing
a crucial information for theories of planetary formation.

\begin{acknowledgements}

We thank Carsten Dominik for providing us with the grid of models of the
physical structure of the studied disks. We also wish to thank E.\ Hugo \& S.\
Schlemmer for providing us the results of their modeling prior to publication.
Support by the Spanish Ministerio de Educaci\'on y Ciencia through project
AYA2004-05792 (A.A.R), the French Projet Nationale PNPS (C.C.) and NSF award
AST-0507421 (M.E.) is gratefully acknowledged.

\end{acknowledgements}

\bibliographystyle{aa}

\end{document}